\documentclass[prl,twocolumn,color,superscriptaddress]{revtex4}
\usepackage{graphicx}
\usepackage{color,amsmath,graphicx,latexsym}
\usepackage{epsfig}
\definecolor{li}{rgb}{0,0,1}
\begin{document}
\title{The route of frustrated 
cuprates from antiferromagnetic  to
ferromagnetic spin-1/2 Heisenberg chains: 
Li$_2$ZrCuO$_4$ \textcolor{black}{as a} missing link near the 
quantum
critical point}

\author{S.-L.\ Drechsler$^{*}$}
\affiliation{Leibniz-Institut f\"{u}r Festk\"{o}rper- und 
Werkstoffforschung (IFW) Dresden, P.O.\ Box 270116, D-01171 Dresden,
Germany}

\author{O.\ Volkova}
\affiliation{Moscow State University, Moscow, 119992, Russia}
\affiliation{Institute of Radiotechnics and Electronics, 
RAS, Moscow 125009, Russia}
\author{A.N.\ Vasiliev}
\affiliation{Moscow State University, Moscow, 119992, Russia}
\author{N.\ Tristan}
\affiliation{Leibniz-Institut f\"{u}r Festk\"{o}rper- und 
Werkstoffforschung Dresden, P.O.\ Box 270116, D-01171 Dresden,
Germany}
\author{J.\ Richter}
\affiliation{Institut f\"ur Theoretische Physik, Otto-
v.\ Guericke-Universit\"at zu Magdeburg, Magdeburg,  Germany}
\author{\protect M.~Schmitt \protect}
\author{H.\ Rosner}
\affiliation{Max-Planck-Institut f\"ur chemische Physik fester Stoffe,
Dresden, Germany}
\author{J.~M\'{a}lek}
\affiliation{Leibniz-Institut f\"{u}r Festk\"{o}rper- und 
Werkstoffforschung
Dresden, P.O.\ Box 270116, D-01171 Dresden, Germany}
\affiliation{Institute of Physics,  
ASCR, Na Slovance 2, CZ-18221 Praha 8, 
Czech Republic}
\author{R.\ Klingeler}
%\author{G.\ Krabbes}
%\author{B.\ B\"uchner}
\affiliation{Leibniz-Institut f\"{u}r Festk\"{o}rper- und 
Werkstoffforschung Dresden, P.O.\ Box 270116, D-01171 Dresden,
Germany}

\author{A.A.\ Zvyagin}
\affiliation{Leibniz-Institut f\"{u}r Festk\"{o}rper- und 
Werkstoffforschung
Dresden, P.O.\ Box 270116, D-01171 Dresden, Germany}
\affiliation{ B.I.\ Verkin Institute for Low Temperature
 Physics \& Engineering, NASU,
% of the Nat.\ Acad.\ of Science of Ukraine, 
 UA-61103 Kharkov, Ukraine}
\author{B.\ B\"uchner}
\affiliation{Leibniz-Institut f\"{u}r Festk\"{o}rper- und 
Werkstoffforschung
Dresden, P.O.\ Box 270116, D-01171 Dresden, Germany}
\date{\today}
\begin{abstract}
\noindent From thermodynamics, LSDA+$U$ studies
%calculations 
and exact 
diagonalizations of a  five-band Hubbard 
%3$d$ 2$p$ 
model on CuO$_2$ stripes
%for finite CuO$_2$ stripes 
we found \textcolor{black}{that}
Li$_2$ZrCuO$_4$ (Li$_2$CuZrO$_4$
in traditional notation)
% s.\Ref.~\onlinecite{remarknotation})
\textcolor{black}{is close} 
to a ferromagnetic critical point.
%$\alpha_c=$ 1/4 within the isotropic spin-1/2 Heisenberg model, where
%$\alpha$ measures the ratio of the antiferromagnetic 
%2nd to the ferromagnetic 1st neighbor 
%exchange integrals $\alpha=-J_2/J_1\approx$ 0.295. 
Analyzing its
%measured 
susceptibility $\chi(T)$ and specific heat $c_p(T,H)$
within a Heisenberg model, we show that 
the ratio of the 2nd to the 1st neighbor
exchange integrals $\alpha$=$-J_2/J_1$$\sim$0.3 is close to
the critical value $\alpha_c$=1/4.  
%Within this model, we 
Comparing with related chain cuprates
we explain the rather 
%unusual 
strong field dependence of $c_p$,
the monotonous down shift of the peak of
$\chi(T)$, and its increase  
for
%zero temperature at 
$\alpha$$ \rightarrow$$ \alpha_c$+0. 
  
%is found close to the predictions of the $J_1$-$J_2$ model.

\end{abstract}
\maketitle

The one-dimensional (1D) spin-1/2 antiferromagnetic (AFM) Heisenberg
model (HM) is one of the most studied many-body models in theoretical
physics. Much of its physics is now well understood based on the
rigorous Bethe-Ansatz method for infinite chains \cite{Johnston} and
on finite cluster calculations.  
%The 
Thermodynamic benchmarks of this model 
relevant here are:
%is characterized by 
(i) single maxima of the spin susceptibility $\chi(T)$ at
$k_{\mbox{\tiny B}}$$T^\chi_{\mbox{\small m}}\approx$ 0.64$J$ and of
the specific heat $c_v(T)$ at $k_{\mbox{\tiny
B}}$$T^{c}_{\mbox{\small m}}\approx$ 0.48$J$,
(ii) 
%a 
%linear
%specific heat 
$c_v \propto T/J$ at $T \rightarrow 0$,
%(ignoring logarithmic and quadratic corrections), 
and (iii) 
%a finite 
$\chi^*(0)=J\chi(0)/Ng^2\mu^2_{\mbox{\tiny B}}=1/\pi^2$ 
%followed by 
and 
%a 
%logarithmically 
%diverging 
%slope of 
d$\chi(T)$/d$T \rightarrow$+$\infty$ 
%(hump)
at $T \rightarrow 0$.
%and (iv) a 
%square-root like behavior of the magnetization at $T=0$:
%$M=g\mu_{\mbox{\tiny B}}(1/2-(1/\pi)\sqrt{H_s-H})$
%at external fields
%approaching the saturation field $g\mu_{\mbox{\tiny B}}H_s=2J$
Hereafter $J\equiv J_1$
 denotes the 
 nearest neighbor (NN) exchange.
%Recently several quasi-1D compounds have been discussed as nearly
%perfect realizations \cite{xx}. 
For ferromagnetic (FM) $J_1 <0$, $\chi(T)\propto 1/T^2$
%diverges 
and $c_{v} \propto \sqrt{T/\mid J_1\mid}$ at $T\rightarrow 0$; $c_v$
shows a broad maximum at $k_{\mbox{\tiny B}}T^{c_v}_{\mbox{\small
m}}=0.35\mid J_1\mid$ and a field induced 2nd maximum at low $T$
%for weak 
and 
%: $
$H <0.008\mid J_1 \mid /g\mu_{\mbox{\tiny B}}$
\cite{junger}. The 
general Hamiltonian $\cal{H}$ with 
%arbitrary
%HM, 
%includes
%ing 
next-nearest neighbors (NNN) $J_2$ 
or further in-chain exchange $J_i$ included
\begin{equation}
{\cal{H}}=\sum_{i} J_1{\mathbf S}_i{\mathbf S}_{i+1}+
J_2{\mathbf S}_i{\mathbf S}_{i+2}+
J_3{\mathbf S}_i{\mathbf S}_{i+3}+... \quad ,
\end{equation}
has also attracted attention 
 due to the frustration caused by AFM $J_2$, 
irrespective of the sign of $J_1$. 
%In the spin-Peierls system GeCuO$_3$, where 
If the $J_i$  
%couplings 
are AFM, the frustration may
cause a spin gap, e.g.\ for $J_2/J_1>0.241$ and
 $J_i=0, i\geq 3$ (adopted mostly below). It strongly supports a
dimerized ground state in spin-Peierls chains such as in GeCuO$_3$
\cite{Castilla}.  Recently, FM-AFM analogs realized 
in most edge-shared chain cuprates  have
caused attention with respect to strong quantum effects \cite{bursill},
to unusual thermodynamics of the disordered
phase \cite{Thanos,Lu,Heidrich}, and to helicoidal ground states found
in some chain cuprates at low $T$ \cite{drechsler,masuda,enderle,gippius,capogna,
drechsler06,choi06,hase,hase2,dmitriev}.  
%Strong quantum effects push
%the chains toward collinear states compared with classical spin chains. 
However, issues like the behavior at
very low $T$ and in magnetic fields near the critical point
$\alpha_c=-J_2/J_1=1/4$
%(for $J_i=0, i\geq 3$) 
are still unclear
%The low-$T$ thermodynamics close to $\alpha_c$ is 
and difficult to study
numerically \cite{Lu} even by the 
transfer matrix renormalization group (TMRG)
method.
For $\alpha > \alpha_c$ the ground state of a classical chain is
formed by a helix with a pitch angle $\phi$ given by $\cos \phi =
-J_1/4J_2\equiv 1/(4\alpha )$. This helix interpolates between a
FM-chain at $0\leq \alpha \leq \alpha_c$ and two decoupled AFM-chains
at $\alpha =\infty$. Noteworthy, $\alpha_c$ is unaffected by quantum
effects \cite{dmitriev}. Since this should hold for the case of
long-range inchain couplings, too, we expect a down(up)shift of
$\alpha_c$ for AFM (FM) $J_i$, ($i \geq 3$):
\begin{equation}
\alpha_c = \frac{0.25}{1+2.25\frac{J_3}{J_2}+4\frac{J_4}{J_2}
+6.25\frac{J_5}{J_2}+9\frac{J_6}{J_2}+...} \quad .
\end{equation}
 Recently, low $T$-$\chi(T)$ data for
Rb(Cs)$_2$Mo$_3$Cu$_2$O$_{12}$ \cite{hase,hase2} have been refitted by
the isotropic $J_1$-$J_2$ HM near $\alpha_c$.  However, both compounds
seem to be affected by Dzyaloshinskii-Moriya  
%type
interactions ${\mathbf D}_{ij}({\mathbf S}_i\times {\mathbf S}_{j}$)
\cite{Lu} and exhibit a very complex, partially unresolved crystal
structure, complicating a theoretical study even more. 

Hence, 
studies of less complex systems described by Eq.\ (1) but with $\mid
\alpha-\alpha_c\mid \ll 1$ are of general interest. Analyzing
$\chi(T)$, $c_P(T,H)$, and the electronic structure of Li$_2$ZrCuO$_4$
we will show that it is a suitable candidate to probe the vicinity of
$\alpha_c$ from the helical side. Together with 
%available 
data for
related 
systems
%chain cuprates 
with $\alpha \geq$ 1 it provides a so far missing link
near $\alpha_c$ to study e.g.\ the $\alpha$-dependence of
relations (i)-(iii), moving from AFM to FM-chains.

\begin{figure}[t]
\includegraphics[width=8.cm,angle=0]{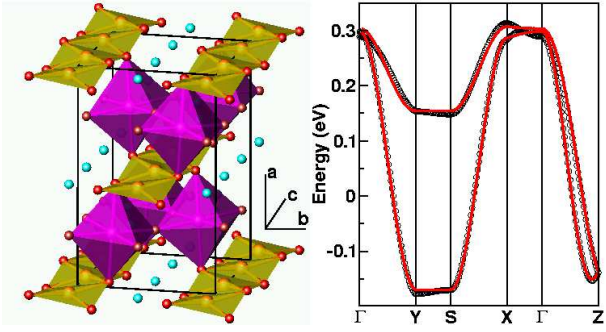}
\caption{(Color online) Crystal and electronic structure 
near the Fermi level $E_{\mbox{\tiny F}}$=0 of 
Li$_2$ZrCuO$_4$.  Left:
Crystal structure, Cu$^{2+}$- large orange $\circ$, 
Zr$^{4+}$- light
magenta
\textcolor{black}{$\circ$}
inside the magenta corner-shared ZrO$_6$ octahedra, red 
\textcolor{black}{$\circ$}- O$^{2-}$, 
and light blue 
\textcolor{black}{$\circ$}- Li$^{+}$ (Li (split) positions 
near Zr 
are omitted for clarity);
the nonplanar edge-shared CuO$_2$ chains (olive-green).
Right: LDA-FPLO band structure ($\circ$) and TB
fit (red line).  
$\Gamma$,X,Y,Z,S are symmetry points in wave vector
notation: (0,0,0);(2$\pi/a$,0,0);(0,$2\pi /b$,0);
(0,0,$2\pi/c$);(2$\pi/a$,2$\pi/b$,0), respectively. 
}
\label{structure}
\end{figure}

The orthorhombic crystal structure of Li$_2$ZrCuO$_4$
\cite{remarknotation} (space group Cccm) 
with the lattice constants $\mathbf a$=9.385~\AA , $\mathbf
b$=5.895~\AA , $\mathbf c$=5.863~\AA \ is shown in Fig.\
\ref{structure}. Here chains (formed by flat edge-shared 
CuO$_4$ tetrahedra like the edge-sharing of CuO$_4$ plaquettes in
other chain cuprates) run along the
$c$-axis.
\begin{figure}[b]
\includegraphics[width=6.0cm,angle=-90]{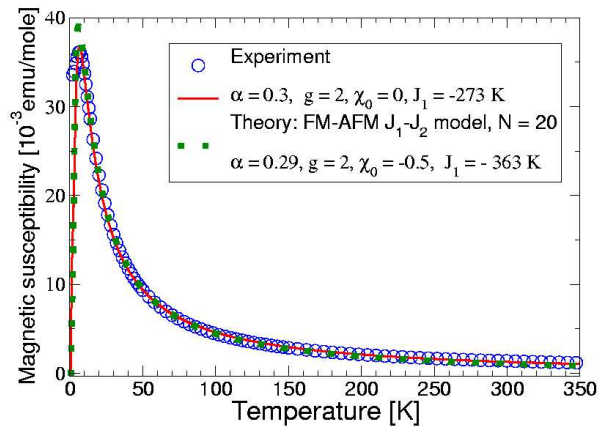}
\caption{(Color online). \textcolor{black}{Magnetic susceptibility 
of Li$_2$ZrCuO$_4$} together with fits
by the $J_1$-$J_2$ model for periodic chains.
% with $N=$20 sites.
}
\label{chi}
\end{figure}
Also the Cu-O bond length of 2.002 \AA \ and the Cu-O-Cu bond angle 
$\gamma=$ 94$^\circ$ resemble those with FM 
$J_1$.

The $\gamma$-polymorph of Li$_2$ZrCuO$_4$ (Li ordered) was
prepared by a solid state reaction of Li$_2$CO$_3$, ZrO$_2$ and CuO
\cite{Dussarrat}. The reagents were mixed in an agate mortar and
fired for a few hours in a Pt boat at 700$^\circ$C to decarbonate
them. Final firing of the pellet
% of low temperature light green
% coloured $\gamma$-Li$_2$ZrCuO$_4$ 
was performed at 1050$^\circ$C for
24 h in a flow of O$_2$ followed by furnace cooling in
O$_2$. Phase purity was confirmed by x ray diffraction.
%Very small amount of CuO impurity was 
%detected by x ray diffraction 
%similarly to results reported in Ref.\ \onlinecite{Dussarrat}.

The magnetization of Li$_2$ZrCuO$_4$ 
%up to 5 T 
measured in a range
2$\leq T \leq$ 350 K for 0.1 T 
by a Quantum Design SQUID magnetometer is
shown in Fig.\ 2. From the observed $T^\chi_{\mbox{\small m}}\approx$
7.6 K one might at first glance expect an AFM spin liquid regime with
%a tiny 
$J_1$ or $J_2\approx$ 12 K, if $\gamma$ is just by chance 
close to that 
%Cu-O-Cu 
bond
angle where $J_1$ changes its sign and either $J_1 \gg J_2>0 $
due to the nonideal chain geometry 
or vice versa 
$J_2 \gg \mid J_1\mid$.  
But the measured $\chi^*(T_{\mbox{\small m}})$ is twice as large
 as the AFM-HM value of 0.1469 
($\chi^{\mbox{\tiny AFM-HM}}(T_{\mbox{\small m}})$=0.0183 emu/mole for $g$=2).
%The analysis of
$1/\chi(T)\propto T+\tilde{\Theta}_{CW}$ reveals a FM Curie-Weiss
temperature $\tilde{\Theta}_{CW}$= -24 K using a  narrow
%highest available
temperature range near 350 K. Both facts exclude any AFM-HM like
scenario. But they point to FM exchange involved in accord with 
fits by the $J_1$-$J_2$ model (Fig.\ 2).
% for Li$_2$ZrCuO$_4$.

Specific heat down to 0.35 K was measured by Quantum Design Physical 
Properties Measurements System (see Fig.\ 3). 
\begin{figure}[t]
\includegraphics[width=6.2cm,angle=-90]{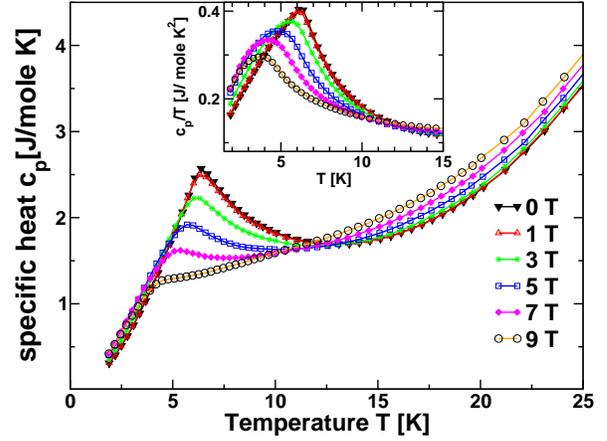}
\caption{\textcolor{black}{(Color online)} Specific heat $c_p$ of Li$_2$ZrCuO$_4$ 
vs.\ \textcolor{black}{$T$} at 
various 
external magnetic fields $H$. 
Inset: the same for $c_p/T$.
} 
\label{heat}
\end{figure}
 %The measured $c_p$
%specific heat 
It shows a relative sharp peak 
%maximum \chi
near 6.4 K at $H$ =0. 
Using $c_p\approx c_v \equiv c$
\cite{remarkcp}, the observed ratio
$T^\chi_{\mbox{\small m}}/T^{c}_{\mbox{\small m}}$= 1.17 differs
from 1.33 predicted by the AFM-HM.
\textcolor{black}{Note that} $T^{c}_{\mbox{\small m}}$ 
nearly coincides
with the $T$ for which d$\chi(T)$/d$T$ becomes 
%At present 
maximal. Hence, it is unclear 
%to what extent it 
\textcolor{black}{whether can this peak be attributed either to}
a $c_p$ anomaly
indicating often
a magnetic phase transition \cite{remarkPT}, or 
\textcolor{black}{to}
a specific feature of 
the disordered phase 
generic for  the 1D
frustrated $J_1$-$J_2$ HM  at 
$\alpha_c < \alpha < 0.4$ \textcolor{black}{.}
Here
%$, near the quantum critical point,$  
$c_{v}$ exhibits a \textcolor{black}{{\it two}}-peak 
structure \cite{Thanos,Lu,Heidrich}:
a sharp peak at low $T$ under consideration and a 
broad one at high 
$T$ hidden in the phonon region 
( $k_{\mbox{\tiny B}}T\sim 0.65\mid J_1\mid \approx$ 260 K in 
the present case).
 Anyhow, with increasing 
field $T^{c}_{\mbox{\small m}}$ is downshifted  and $c_p(T_{m})$ 
is suppressed but  
$c_{p}(T)$ increases
rapidly for $T\geq$ 12K, well above a possible phase 
transition near 6K.
  
 A similar strong $H$-dependence 
is found in full diagonalization studies of large rings, 
where the low-$T$ peak is first downshifted with increasing $H$ and upshifted
at higher $H$ (Fig.\ 4).
The strong $H$-dependencies of 
both
$\Delta c(H)$=$c(H)$-$c(0)$ and -$\Delta T^{c}_{\mbox{\small m}}$=
$T^{c}_{\mbox{\small m}}$(0)-$T^{c}_{\mbox{\small m}}(H) \propto H^2$
already at weak 
fields $H \leq $ 9T 
result from the vicinity to $\alpha_c$ \cite{remh}.
Adding a usual lattice contribution 
$c_{lat}\propto T^3$ ($\Theta_{D}$=220 K)
to the calculated spin specific heat
within the isotropic $J_1$-$J_2$ HM
the experimental data are best described by  $\alpha =0.3$ (Fig.\ 4). 
\begin{figure}[t]
\begin{center}
\includegraphics[width=6.8cm,angle=0]{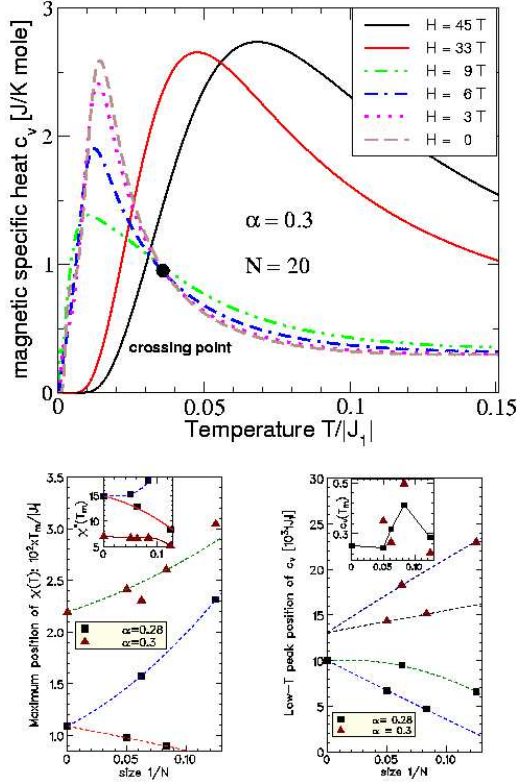}
\end{center}
\vspace{-0.6cm}
\caption[t]{(Color online) Specific heat $c_v$ vs.\ 
temperature in units of 
$\mid J_1 \mid $ for various magnetic fields $H$ within 
the FM-AFM-$J_1$-$J_2$ HM for $\alpha=-J_2/J_1$=0.3 
and a ring with $N$=20 sites
(upper row).
Finite size effects  for the maximum of $\chi(T)$ (left) 
and the low-$T$ maximum of $c_v(T)$ (right)
(lower row). The $N=\infty$ values are taken from
TMRG results of Ref.\ \onlinecite{Lu}. 
}
 \label{band}
\end{figure} 
From the low $H$-crossing point near 12 K 
%and the lower peak (extrapolated to $N=\infty$)
% near 0.0131 $\mid J_1\mid$ for $H$=0, 
we estimate $J_1\approx$ 405 K to $\sim$
 %in accord with
 363 K 
%derived from 
using 
the $\chi(T)$ data 
%but 
for 
%a slightly smaller $J$ ratio
$\alpha = 0.29$. The low-$T$ peaks 
extrapolated to $N=\infty$
would be expected near
 $k_{\mbox{\tiny B}}T^{c}_{\mbox{\small m}}\approx 0.013(0.0115)\mid J_1\mid$,
respectively, 
i.e.\ near 5.3(4.2) K below the observed one at 6.4 K, similarly 
as the expected $T^\chi_{\mbox{\small m}}\approx$ 4.5$\pm$ 1.7 K is below 
the observed one near 7.6  K (Figs.\ 2-4).
\textcolor{black}{These observations are in accord with the 
scenario
of a 
phase transition at 6 K  as discussed above.}
%Hence, the  \textcolor{black}{emergence of such a} low-$T$ peak 
%for $c_p$ \textcolor{black}{generic for}  the spin liquid phase
%might be \textcolor{black}{simply prevented}  
%by a phase transition \textcolor{black}{with a $c_p$-anomaly} 
%\textcolor{red}{of similar shape}.  
The slightly different results 
%$\alpha$ and $J$ values
from \textcolor{black}{fitting $\chi (T)$ or $c_p(H,T)$}  
might be due to  anisotropies 
%e.g.\ a DM-term caused by the low symmetry of our
%nonideal chains 
and  
%due to 
interchain coupling.
% as well as from remaining finite size effects. 

To estimate 
%the strength of 
the 
\textcolor{black}{interchain coupling}, we consider
the measured Curie-Weiss temperature 
$\tilde{\Theta}_{CW}=r\Theta_{CW}$= -24 K, 
where $1\geq r(T)\approx$ 0.25 is estimated from the
calculated $d\chi^{-1}(T)$/d$T$ taken at the highest available 
$T$. Here it is  still outside  the asymptotical CW-range 
%due to missing high-$T$ data for a 
$k_{\mbox{\tiny B}}T\gg \mid J_1 \mid $, where $r\rightarrow 1$.
The high-$T$ expansion of $\chi(T)$ yields
$\Theta_{CW}=0.25\sum_i z_iJ_i$, i.\ e.\
%can be rewriten as
%we estimate
%the 
%strength of the 
%frustrated 
%interchain coupling 
%$J_{\perp}$ from
\begin{equation}
2\tilde{\Theta}_{CW}/r=J_1\left(1-\alpha \right)  
+J_3+J_{\perp}+2J_{d1}+2J_{d2} \quad , 
%\nonumber
\end{equation}
where the neighbor number 
$z_i=$ 2 for couplings along the $c$ and $b$-axes and 
$z_i$=4 for  diagonal interchain exchange (d1,d2) within the $b,c$-plane. 
From the tight binding (TB) 
%analysis 
fit of the band dispersion  
we find similar direct and diagonal interchain transfer integrals
$t_{\perp}\approx t_{d1}\approx t_{d2}$.  
\textcolor{black}{Setting}
$J_\perp=J_{d1}=J_{d2}$, we found $ \mid J_1 \mid$, $J_2 \gg
J_{\perp}\approx $ 9 K in accord with the LDA results
($J_{\perp}$$\approx$ 7 K).  Thus, the 
\textcolor{black}{adopted} 1D magnetic 
approach is \textcolor{black}{a reasonable starting point}
despite the more 2D electronic structure seen e.g.\
along the symmetry lines $\Gamma$-Y and S-X in Fig.~1.

To get insight into the $J$-set obtained above, we 
%also 
performed
calculations of the electronic and magnetic structure within the local
(spin) density approximation (L(S)DA). In addition, LSDA+$U$
calculations and exact diagonalizations for an appropriate extended
multiband Hubbard model were carried out to take the strong
correlation for the Cu 3$d$ holes into account.
The LDA calculations (Perdew-Wang92 parametrization) were performed
using the full-potential local-orbital minimum-basis scheme 
(FPLO, vers. 5.00-19)\cite{koepernik}. We employed a basis set of
Cu(3$s$3$p$):(4$s$4$p$3$d$), O(2$s$2$p$3$d$),
Zr(4$s$4$p$):(5$s$5$p$4$d$), and Li(1$s$):(2$s$2$p$3$d$).
For the LSDA+$U$ in the AFM version \cite{czyzyk94} we used
$U_{3d}$=6.5$\pm$1.5 eV and the intraatomic 
exchange $J_{ex}$=1 eV. Following the approach of
analyzing total energy differences for various magnetic
superstructures \cite{nitzsche}, we obtain $J_1$= -151$\mp35$ K and
$J_2$=35$\mp$12 K.
Using a typical 
%value for the 
one-band Hubbard $U_{eff} \approx$ 3.5 eV as well as
%our NN-transfer integrals 
$t_2$ and $t_3$ from the TB fit of the
%LDA-
band
%s 
at $E_{\mbox{\tiny F}}$ (Fig.\ 1) results in $J_2$=46 K and
$J_3$=6 K employing $J_i=4t_i^2/U$.
Thus, we arrive close to $\alpha_c$=0.195 in the present case of
$J_3\neq 0$ (s.\ Eq.~(2)).

Finally, a collection of known $T^{\chi}_{\mbox{\small m}}/J_2$
and $\chi(T_{\mbox{\small m}})$ values
from other chain cuprates we
derived from their $\chi(T)$ data 
\cite{drechsler,enderle,gippius,drechsler06,baran}, is shown in 
Fig.\ 5 \cite{remarkplot}.
\begin{figure}[b]
\begin{center}
\includegraphics[width=9.0cm,angle=0]{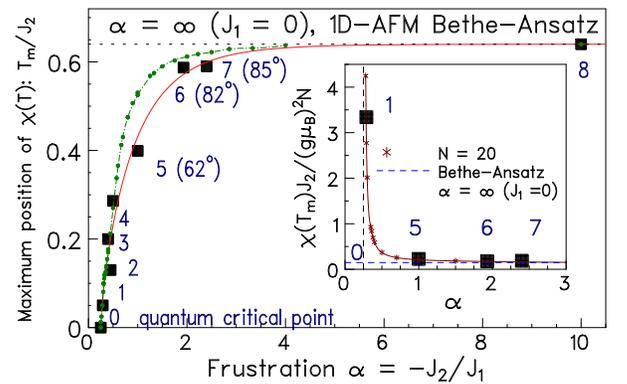}
\end{center}
\vspace{-1.cm}
\caption[]{\textcolor{black}{(Color online)} 
Empirical $T^\chi_{\mbox{\small m}} $ in units of the
fitted
%NNN exchange 
$J_2$ value 
of the FM-AFM $J_1$-$J_2$-model
%vs.\ 
%the  $\alpha=-J_2/J_1$ 
for several frustrated chain cuprates (black squares): 0- $\alpha_c$=1/4, 
%the FM critical point, 
%$\alpha = \infty$ corresponds to two decoupled AFM chains. 
1- Li$_2$ZrCuO$_4$, 
2- Pb$_2$[CuSO$_4$(OH)$_2$],
% \cite{baran},
3- Rb$_2$Cu$_2$Mo$_3$O$_{12}$, 
4- Cs$_2$Cu$_2$Mo$_3$O$_{12}$, 5- LiCu$_2$O$_2$, 6- NaCu$_2$O$_2$,
7- LiVCuO$_4$,  and 8- SrCuO$_2$. The measured pitch 
%angle 
is given in 
brackets.    
The filled green $\circ$ denote the full diagonalization 
results of the $J_1$-$J_2$ model on 
rings with $N$=20 sites. Inset: the maximum value of $\chi(T)$.
}
 \label{chifrust}
\end{figure} 
In particular, it is clear
why the large-$\alpha$ chains in SrCuO$_2$ and
LiVCuO$_4$   are often
regarded as  AFM-HM archetypes \cite{vasiliev}.
Only after the discovery of spirals,
detailed inelastic neutron scattering
studies, and
our three component theoretical analysis
(HM, Cu-O Hubbard model, LDA)
 initial assignments for LiVCuO$_4$ and LiCu$_2$O$_2$
were corrected \cite{masuda,enderle,gippius,drechsler}.
Similarly, among systems assigned so far as ''perfect'' 
realizations
of the AFM/FM HM (e.g.\ \cite{remarkPT,chagas})
could be further $J_1$-$J_2$ candidates worth to be 
revisited. 
Similar plots which accent the FM critical point 
can be made also for the low-$T$ 
peak of $c_v$ or $\chi(0)$ (which monotonously increases and 
diverges finally as $\alpha \rightarrow \alpha_c$) \cite{remarkchi}.
We expect that \textcolor{black}{a vanishing}
$T^{\chi (c)}_{\mbox{\small m}}$ 
%\rightarrow$ 0 
and \textcolor{black}{a diverging} 
$\chi(T)$
%\rightarrow \infty$ 
for $T \rightarrow$ 0
\textcolor{black}{in}
 approaching $\alpha_c$ \textcolor{black}{are} generic for a FM
critical point.
 It should hold also 
for models beyond the $J_1$-$J_2$ HM. 
Further couplings
do affect the helical phase in changing e.g.\ 
the pitch and
%. In the present context their main effect is
%to change   
$\alpha_c$  (see Eq.\ (2)). 

Comparing 
%the obtained 
$\alpha \sim \alpha_c$ for Li$_2$ZrCuO$_4$ 
with $\alpha \gg \alpha_c$
we found for other chain cuprates
the question arises, what is the 
microscopic reason for? There are at least two options:
 (i) an enhanced 
$\mid J_1 \mid$ at a standard
$J_2$ value and vice versa (ii) a slightly enhanced 
$\mid J_1 \mid$
at a reduced $J_2$. Case (i)
can be ascribed to 
enhanced FM contributions to $J_1$ which arise from the 
direct exchange $K_{pd}$  or from 
the Hund's rule coupling at the sharing
O ions within a five-band 
Cu 3$d$ O $2p$ 
extended Hubbard model.
Unfortunately, there is no 
generally accepted  $K_{pd}$ value; but  
it is 
%\textcolor{black}{there is} 
%consensus 
%that it is 
the most sensitive
quantity for the determination of $J_1$
in edge-shared cuprates \cite{braden}. Nevertheless,   
usually $K_{pd}$ is treated as
a fit parameter:
% \cite{braden,mizuno}. 
%The 
%experimental data 
%$\chi(T)$, optical conductivity and O 1$s$ XAS spectral data 
%of  
The well-studied Li$_2$CuO$_2$
can be described with $K_{pd}$=50 meV \cite{mizuno}, 
whereas 
microscopic
%quantum chemistry 
calculations for 
%the high-$T_c$ parent compound 
La$_2$CuO$_4$ yield 180 meV \cite{Hybertsen}
and a structural analysis 
of GeCuO$_3$ was performed adopting 
$K_{pd}$=110 meV \cite{braden}. 
Within the Cu 3$d_{yz}$ O 2$p_{y},p_{z}$ 
extended Hubbard model 
%applied to 
for planar 
%Cu$_n$O$_{2n}$  and 
Cu$_n$O$_{2n+1}$ open chains
($n \leq 5$),
we adopted $K_{pd}$=70 meV. From a direct mapping onto the 
$J_1$-$J_2$ HM using 
%$U_d=$8.5eV, $U_p$=4.4eV, 
two O site energies 
$\Delta_{p_yd}$=2.7eV and 
%(-0.5) eV, 
$\Delta_{p_zd}$= 3.2eV as well as 
%standard fixed 
Li$_2$CuO$_2$\textcolor{black}{-like}
parameters, we found 
  $J_1$=-317  K, $J_2$= 90 K, 
and $\alpha=$0.284,  
close to our empirical values. 
In 
%the opposite 
case (ii) 
%of considerably reduced $J_2$ 
supported by the LSDA+$U$ results,
we arrive also close to $\alpha_c$. 
%for $U=5.5$(6.0) eV at $J_1$=-17.2(-15.6) meV, 
%$J_2$=4.6 (4.1) meV, $J_3$=0.5 meV, respectively,
%which are also close to 
%the quantum critical point, 
Here $J_2$ amounts 46 K, only.
%Hopefully, 
From a comparison 
with other 
%related Rb(Cs)$_2$Cu$_2$Mo$_3$O$_{12}$ 
cuprates in Fig.\ 5 more insight
will be gained
into the nature of the exchange in cuprates 
and  
%nature 
into the novel physics generated by a quantum 
FM critical point.
% as a central topic in modern solid state physics.

%about
%50 to 70\% 
%60\% of
%the former. 

%of the $J$'s from the  set.
%Further work is required 
%to find refined parameters and 
%to resolve this 
%puzzle. 

To conclude, we have shown that a growing number of
%several 
%edge-shared chain cuprates 
edge-shared chain cuprates form 
a special family 
%interesting subclass 
which thermodynamics 
can described  within the 
%isotropic 
$J_1$-$J_2$ model with FM NN and AFM NNN exchange. 
%as a starting point. 
Moving
from the AFM-HM 
%approaching  
towards
%the vicinity of 
$\alpha_c$, 
%the FM critical point 
%to ferromagnetism have been observed for 
almost achieved for
Li$_2$ZrCuO$_4$, observed monotonic  changes  can be 
explained.
Only chains near the FM critical point show peculiar
%thermodynamic 
physical
properties such as the strong
$H$-dependence of $c_p$ in a large $T$ range
reported here.
% for Li$_2$ZrCuO$_4$.
Further studies of Li$_2$ZrCuO$_4$ 
%remarkable compound
%challenging
%chain cuprate 
at very low $T$, under pressure, and in 
high fields  
are highly desirable. \textcolor{black}
{If the observed $c_p$-peak 
%near 6 K
is related to magnetic ordering, neutron diffraction 
below 6 K should reveal 
a spiral with a pitch 
%well 
below the minimum
value of
62$^\circ$ observed so far  among 
edge-shared chain cuprates for
 LiCu$_2$O$_2$ \cite{masuda}.
Inelastic neutron scattering studies 
%measurements 
%on single crystals 
might 
be helpful to 
refine
%values of 
the 
%involved 
exchange integrals, especially 
with respect to the interchain coupling.}
% responsible for the phase transition.}

%\begin{acknowledgment}
 
We thank  
H.\ Eschrig, D.\ Dmitriev, V.\ Krivnov, M.\ Hase and R.\ Kuzian, for 
%helpful and 
discussions and 
N.\ Wizent for experimental support.
 Support from grants 
\textcolor{black}{(DFG,  
(486 RUS 113/864/0-1, (B), 
KL 1824/2, (T,K), (M), the 
Emmy-Noether program (S,R)),
GIF,(R), CRDF, RU-P1-2599-MO-04, 
and RFBR 06-02-16088 (V,V))
is} 
%gratefully 
acknowledged.
%\end{acknowledgments}


\begin{thebibliography}{99}
\bibitem{remarknotation}We use Li$_2$ZrCuO$_4$ and GeCuO$_3$ instead of 
Li$_2$CuZrO$_4$ \cite{Dussarrat} and CuGeO$_3$
 and analogously for other 
edge-shared systems
to accent \textcolor{black}{their cuprate 
(not zirconate or germanate, etc.)} nature. 
A more precise, but somewhat lengthy, notation would be
Li$_2$ZrO$_2$CuO$_2$ and GeOCuO$_2$ to show the inert cationic  
and the active magnetic anionic cuprate complexes.
\bibitem{remarkcp}
%We ignore the \textcolor{black}{small} 
%difference between $c_p$ and $c_v$. 
%%For solids it is usually
%% very small.
%: xxxxxx
%To be added by R. Klingeler. 
\textcolor{black}{Here the indices are not dropped
because} $c_v$ denotes also the 
%indices 
%distinguish
%to help the reader to distinguish 
calculated and $c_p$ the 
measured values.
\bibitem{remarkPT}
For 1 K $\leq T\leq 4$ K we observed a
 $c_p\propto T^{1.8}$-law 
 and a linear law for  0.35 K $\leq T \leq $ 1 K.
Its relative large Sommerfeld constant 
$\gamma \approx 90$ mJ/(moleK$^2$) is noteworthy.
 The exponent 1.8 of the 
former range, being in between a 3D FM (3/2) and 
a quasi-2D N\'eel (2) state,
resembles those of quasi-1D 
FM Cu-peptides (1.73 and 1.84) \cite{chagas}.
\bibitem{remh}
\textcolor{black}{In other chain cuprates  with comparable 
or even smaller $J$-values
%smaller magnitudes
%of the exchange integrals 
but with 
$\alpha \sim $1,
considerably higher fields $H \sim$ 20-30 T 
are predicted to affect $c_{v}$ in the para phase
(aside from  special effects caused by 
%the presence of
anisotropic 
%exchange 
terms in $\cal{H}$).}
\bibitem{remarkplot}
For $\alpha$=0.5 and $N=20$ $\chi(T)$ shows two maxima.
In the plot the upper broad one is shown. 
The 
\textcolor{black}{roughly estimated} points 3 and 4 
%for Rb(Cs)$_2$Cu$_2$Mo$_3$O$_{12}$ 
%are obtained 
\textcolor{black}{follow}
from Eq.\ (3) using the $J_1$, 
$T^{\chi}_{\mbox{\small m}}$,   
$\tilde{\Theta}_{CW}$ reported \cite{hase}, 
$r$=0.5(0.9), and adopting 
$J_{\perp}/J_1=\mp$0.2 as well as $z_i=1$ for the interchain
interaction. 
However, 
with respect to  mentioned uncertainties for  
Rb(Cs)$_2$Cu$_2$Mo$_3$O$_{12}$ 
they
should be 
regarded as rough estimates.
\bibitem{remarkchi}At $\alpha_c$ 
%according to our cluster calculations ($N$=20)
% 
the divergency of $\chi(T)$ at $T \rightarrow 0$
%\propto
%1/T^{1+\beta}$, where $\beta \approx $0.12233,  
is certainly weaker 
%smaller 
than for  ideal FM chains at $\alpha =0$.
\end{thebibliography}
\end{document}